# Effects of Nanoparticle Geometry and Size Distribution on Diffusion Impedance of Battery Electrodes

Juhyun Song[a], and Martin Z. Bazant[a,b]

[a]Department of Chemical Engineering and [b]Department of Mathematics,

Massachusetts Institute of Technology,

77 Massachusetts Avenue, Cambridge, Massachusetts 02139, United States

bazant@mit.edu

**Abstract**

The short diffusion lengths in insertion battery nanoparticles render the capacitive behavior of bounded diffusion, which is rarely observable with conventional larger particles, now accessible to impedance measurements. Coupled with improved geometrical characterization, this presents an opportunity to measure solid diffusion more accurately than the traditional approach of fitting Warburg circuit elements, by properly taking into account the particle geometry and size distribution. We revisit bounded diffusion impedance models and incorporate them into an overall impedance model for different electrode configurations. The theoretical models are then applied to experimental data of a silicon nanowire electrode to show the effects of including the actual nanowire geometry and radius distribution in interpreting the impedance data. From these results, we show that it is essential to account for the particle shape and size distribution to correctly interpret impedance data for battery electrodes. Conversely, it is also possible to solve the inverse problem and use the theoretical "impedance image" to infer the nanoparticle shape and/or size distribution, in some cases, more accurately than by direct image analysis. This capability could be useful, for example, in detecting battery degradation *in situ* by simple electrical measurements, without the need for any imaging.

**Introduction**

In impedance spectra of intercalation battery electrodes, the response at low frequencies corresponds to solid-state transport of charge carriers (*e.g.* lithium ions and electrons in lithium ion batteries) in the active material. The transport of charge carriers is dominated by ionic diffusion in many battery materials due to the high mobility of electrons [1, 2]. For traditional battery electrodes with large particle sizes, the Warburg-type diffusion impedance, which draws a $45°$ line in the complex plane representation (Nyquist plot), has been widely reported at low frequencies. Such response is well-described by a linearized diffusion model in a semi-infinite planar domain. This model leads to the original Warburg impedance formula, $Z_W = A_W (1-i) \omega^{-1/2}$, where $A_W$ is the Warburg coefficient, $i = \sqrt{-1}$ is the unit imaginary number, and $\omega$ is the radial frequency [3]. With conventional particle sizes in micron scale or larger, the diffusion penetration depth from the active material/electrolyte interface does not effectively reach the center of an intercalation particle in the typical frequency window (MHz ~ mHz) of an impedance measurement, and the original Warburg impedance could be widely used in interpreting diffusion impedance of battery electrodes [4].

However, impedance spectra of modern thin film and nanoparticle battery electrodes show a distinguished feature in the diffusion impedance: the response transitions from the original Warburg behavior to a capacitive behavior in a lower frequency range, represented by a vertical line in the Nyquist plot [5-10]. This transition is observed because the diffusion penetration depth can reach the impermeable current collector of a thin film electrode or the reflective center of a nanoparticle at accessible low frequencies, due to short diffusion lengths in the thin film and nanoparticles. In the lower frequency range, the sinusoidal stimuli in impedance spectroscopy lead to effectively filling up and emptying the active material, much like a capacitor, resulting in the vertical capacitive behavior. When diffusion impedance has such behavior due to bounded diffusion space in active material, it is referred to as bounded diffusion (BD) impedance throughout this article, while it has been inconsistently called by various other names [2], such as open-circuit (blocked) diffusion [1], finite-space Warburg [4, 11], and diffusion impedance with impermeable [12] or reflecting [13] boundary conditions.

The BD impedance has different properties depending on electrode configuration in terms of diffusion geometry and length distribution in active material. Ionic diffusivity can be obtained from the diffusion impedance, provided the configuration factors from modern electron microscopy and the functional formula from a mathematical model [14]. While theoretical formulae of the BD impedance have been derived for a thin film electrode and nanoparticle electrodes with some model particle geometries [2, 6, 12, 13, 15, 16], they have not been widely applied by experimentalists. In most applications, only the original Warburg impedance model and an one-dimensional BD impedance model have been exclusively used without considering the actual curved particle shape and particle size distribution in interpreting diffusion impedance [7-10, 17-19]. Likewise, only these two models are built into commercial data-processing software products (*e.g.* Zview from Scribner Associates, Inc., ZSimpWin from EChem Software, and Echem Analyst from Gamry Instruments).

In this article, we reformulate BD impedance models for planar, cylindrical, and spherical diffusion geometries and incorporate them into an overall impedance model of battery electrodes, investigating the effect of particle geometry and size distribution on diffusion impedance. The models assume that the active material forms a solid solution of intercalated ions, which has isotropic transport properties and high electron mobility. (Other factors affecting impedance, such as phase separation, crystal anisotropy and low electron mobility, are beyond the scope of this article, but are currently under investigation by our group.) Various versions of the model were applied to experimental impedance data of a silicon nanowire electrode, which provides an ideal test case to study the effects of including the actual nanowire geometry and radius distribution in impedance models. Through this application, we show that it is essential to account for particle geometry as well as size distribution to accurately interpret impedance spectra of battery electrodes.

### Theoretical Model

In impedance spectroscopy, a small sinusoidal stimulus either in potential or current is applied about a reference state, and other variables are perturbed accordingly. Each relevant variable can be written as a superimposition of two terms: a term describing the reference state response in the absence of the perturbation, and another term describing the perturbation about the reference state. The small amplitudes of the perturbations allow mathematical linearization of the system, and thus the perturbations in all relevant variables have a sinusoidal form with an identical frequency. An arbitrary system variable, $X$, can be expressed as

$$X = X_{ref} + \text{Re}\left[\hat{X}e^{i\omega t}\right] \quad (1)$$

where $i$ is the unit imaginary number, $\omega$ is the radial frequency, and $t$ is the time variable. The former term, $X_{ref}$, represents the reference state response, and the latter term represents the sinusoidal perturbation in $X$ with a complex exponential, $e^{i\omega t}$, and the Fourier coefficient, $\hat{X}$. The Fourier transformation of the perturbation yields $\hat{X}$, which is a complex number containing information related to the magnitude and the phase of the perturbation. $X$ can be either current density, $j$, potential, $\phi$, or local concentration, $c$.

*Bounded diffusion impedance* − The system under initial investigation is a thin film or a single nanoparticle of active material, in which ions intercalate from its interface with electrolyte, and electrons come from its interface with a current collector or a conducting agent. We take an equilibrium reference state with uniform concentrations of the charge carriers, restricting our model to materials that form a solid solution with the ions. In most active materials, the mobility of electrons is much higher than that of ions [2]. As electrons become freely available in the system, the mean electric field quickly relaxes, resulting in local charge neutrality in the bulk [11, 20]. Under such conditions ionic diffusion dominates transport of the charge carriers, and a neutral diffusion equation, Fick's law, can be recovered for the ion material balance in the system [21].

$$\frac{\partial c}{\partial t} = D_{ch}\nabla^2 c \quad (2)$$

where $c$ is the ion concentration, $t$ is the time variable, and $D_{ch}$ is the chemical diffusivity of ions in the active material. Impedance behavior in a system with comparable electron and ion mobilities has been studied by several groups [21-23].

We hereby focus on model electrode configurations, including a thin film electrode, and nanoparticle electrodes with planar, cylindrical, and spherical particles. Figure 1 shows the model electrode configurations and the corresponding solid-state diffusion geometries in the active material. Under the assumption of isotropic transport properties, the diffusion equation can be reduced to an ordinary differential equation in frequency-space domain through Fourier transformation.

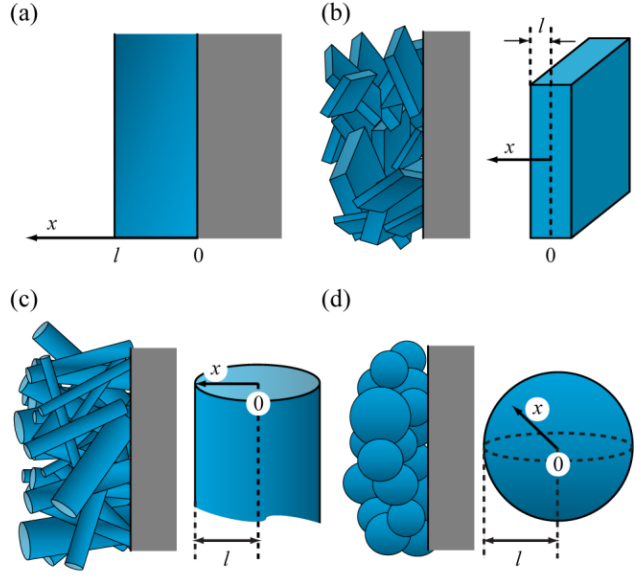

**Figure 1.** Model electrode configurations, particle geometries, and corresponding coordinate systems, where the blue region and the gray region represent the active material and the current collector, respectively: (a) thin film electrode, (b) electrode with planar particles, (c) electrode with cylindrical particles, and (d) electrode with sphere particles

$$i\omega \hat{c} = D_{ch}\nabla^2 \hat{c} = \frac{D_{ch}}{x^{n-1}}\frac{d}{dx}\left(x^{n-1}\frac{d\hat{c}}{dx}\right) \qquad (3)$$

where the spatial variable, $x$, is the distance from the current collector in the thin film, or the distance from the center of symmetry in the planar, cylindrical, and spherical nanoparticles (see Figure 1). The derivative term accounts for variation in the $x$-normal area with respect to $x$, where $n$ is the dimension number: 1 for a thin film electrode and a planar nanoparticle, and 2 and 3 for cylindrical and spherical nanoparticles, respectively.

One of the boundary conditions describes the impermeability of ions at the current collector of a thin film electrode or the symmetry at the center of a nanoparticle.

$$\left.\frac{d\hat{c}}{dx}\right|_{x=0} = 0 \qquad (4)$$

This impermeable or reflective boundary condition indicates that the diffusion space is bounded. The other boundary condition applies Faraday's law at the active material/electrolyte interface to correlate the ion flux and the intercalation current density.

$$\hat{j}_{intc} = -eD_{ch} \left.\frac{d\hat{c}}{dx}\right|_{x=l} \qquad (5)$$

where $j_{intc}$ is the intercalation current density, $e$ is the electron charge constant, and $l$ is the film thickness in a thin film, a half of the thickness in a planar nanoparticle, or the radius in a cylindrical and a spherical nanoparticle. With these boundary conditions, the differential equation can be integrated to give the perturbation profile of ion concentration in the model geometries.

The contribution of solid-state diffusion in the active material appears in an equilibrium potential of the intercalation reaction, since it is a function of the ion concentration at the surface where the reaction takes place. Therefore, the definition of local diffusion impedance takes a partial derivative of the equilibrium potential with respect to ion concentration, and the perturbation in ion concentration at the surface, in place of a potential perturbation [6, 12, 15].

$$z_D = \frac{\Delta\hat{\phi}_{eq}}{\hat{j}_{intc}} = \left(\frac{\partial \Delta\phi_{eq}}{\partial c}\right)\frac{\hat{c}|_{x=l}}{\hat{j}_{intc}} \qquad (6)$$

where $\Delta\phi_{eq}$ is the equilibrium potential of the intercalation reaction, and $z_D$ is the local diffusion impedance. This definition is applied to a system of bounded diffusion space to define local BD impedance.

Properties of the BD impedance can be well-studied when the equations are reduced to their dimensionless forms by proper scaling. The frequency can be scaled by the diffusion characteristic frequency, $\omega_D = D_{ch}/l^2$, that appears when non-dimensionalizing the diffusion equation.

$$\tilde{\omega} = \frac{\omega}{\omega_D} \qquad (7)$$

As the diffusion characteristic frequency is approached, the diffusion penetration depth reaches the impermeable current collector of a thin film electrode or the symmetric center of a particle. The local BD impedance can be scaled by the BD impedance coefficient, $\rho_D = \left(-\partial\Delta\phi_{eq}/\partial c\right)\left(l/eD_{ch}\right)$, which becomes its apparent scale when Equations (5) and (6) are combined.

$$\tilde{z}_D = \frac{z_D}{\rho_D} \tag{8}$$

In Table 1, dimensionless forms of the local BD impedance for the model diffusion geometries are summarized with their asymptotic behaviors. $\tilde{z}_{D\infty}$ and $\tilde{z}_{D0}$ are the asymptotic approximations of $\tilde{z}_D$ at high and low frequencies, respectively [6, 12, 24].

*Local interface impedance* – Local electroactive processes on the active material/electrolyte interface include charging of the double layer and intercalation of ions into the active material. The electrochemical double layer develops at the interface due to the potential drop across it. The double layer charging process can be modeled with the ideal capacitor equation. Using Fourier-transformed variables, the current density becomes

$$\hat{j}_{dl} = i\omega \hat{q}_{dl} = i\omega C_{dl} \Delta \hat{\phi} \tag{9}$$

where $\hat{j}_{dl}$ is the double layer charging current density, $q_{dl}$ is the double layer charge, $C_{dl}$ is the double layer capacitance, and $\Delta \phi$ is the potential drop across the active material/electrolyte interface.

**Table 1. Dimensionless local bounded diffusion impedance and their asymptotic approximations for the model electrode configurations and particle geometries**

| | Thin film electrode and planar particle ($n=1$) | Cylindrical particle ($n=2$) | Spherical particle ($n=3$) |
|---|---|---|---|
| $\tilde{z}_D$ | $\dfrac{\coth\left(\sqrt{i\tilde{\omega}}\right)}{\sqrt{i\tilde{\omega}}}$ | $\dfrac{I_0\left(\sqrt{i\tilde{\omega}}\right)}{\sqrt{i\tilde{\omega}}I_1\left(\sqrt{i\tilde{\omega}}\right)}$ | $\dfrac{\tanh\left(\sqrt{i\tilde{\omega}}\right)}{\sqrt{i\tilde{\omega}} - \tanh\left(\sqrt{i\tilde{\omega}}\right)}$ |
| $\tilde{z}_{D\infty}$ ($\tilde{\omega} \gg 1$) | $\dfrac{1}{\sqrt{2\tilde{\omega}}}(1-i)$ | $\dfrac{1}{\sqrt{2\tilde{\omega}}}(1-i) - \dfrac{1}{2\tilde{\omega}}i$ | $\dfrac{1}{\sqrt{2\tilde{\omega}}}(1-i) - \dfrac{1}{\tilde{\omega}}i$ |
| $\tilde{z}_{D0}$ ($\tilde{\omega} \ll 1$) | $\dfrac{1}{3} - \dfrac{1}{\tilde{\omega}}i$ | $\dfrac{1}{4} - \dfrac{2}{\tilde{\omega}}i$ | $\dfrac{1}{5} - \dfrac{3}{\tilde{\omega}}i$ |

Another current contribution comes from intercalation of ions into the active material. The intercalation kinetics can be modeled with the Butler-Volmer equation which, in general, describes charge transfer reaction rates. The intercalation current density can then be written as

$$j_{intc} = j_0 \left[ \exp\left(\alpha \frac{e\eta}{kT}\right) - \exp\left((\alpha-1)\frac{e\eta}{kT}\right) \right] \tag{10}$$

where $\alpha$ is the electron transfer symmetry factor ($0 < \alpha < 1$), $j_0$ is the exchange current density, and $\eta = \Delta\phi - \Delta\phi_{eq}$ is the surface overpotential. Linearization of the Butler-Volmer equation should consider that both $j_0$ and $\Delta\phi_{eq}$ fluctuate due to the perturbation in ion concentration at the active material surface. When the system is perturbed around an equilibrium reference state, the two exponential terms evaluated at the reference state cancel each other out, and thus the perturbation in $j_0$ does not effectively contribute to the impedance response. On the other hand, the perturbation in $\Delta\phi_{eq}$ brings the contribution from solid-state diffusion in the active material, and introduces the local diffusion impedance. Taking the Fourier transformation, the linearized Butler-Volmer equation becomes

$$\begin{aligned}\hat{j}_{intc} &= \frac{j_0 e}{kT}\left(\Delta\hat{\phi} - \left(\frac{\partial \Delta\phi_{eq}}{\partial c}\right)\hat{c}\Big|_{x=l}\right) \\ &= \frac{1}{\rho_{ct}}\left(\Delta\hat{\phi} - z_D \hat{j}_{intc}\right)\end{aligned} \tag{11}$$

where $\rho_{ct} = kT/j_0 e$ is the charge transfer resistance, and the definition of local diffusion impedance was used. The equation can be rearranged to give a generalized Ohm's law, which leads to a circuit analog of the ion intercalation process.

$$\left(\rho_{ct} + z_D\right)\hat{j}_{intc} = \Delta\hat{\phi} \tag{12}$$

This indicates that ion intercalation could be represented by a series circuit of $\rho_{ct}$ and $z_D$, given a small perturbation.

We assume an independent parallel arrangement of the double layer charging current and the intercalation current on the active material/electrolyte surface, following Randle and Graham [25, 26].

$$\hat{j}_{tot} = \hat{j}_{dl} + \hat{j}_{intc} \tag{13}$$

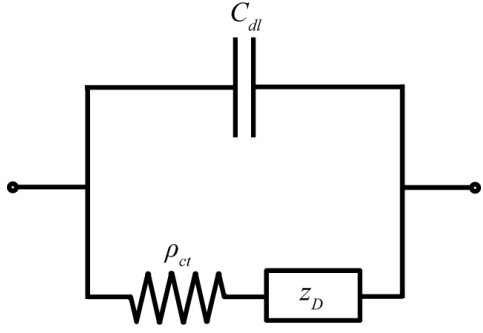

**Figure 2.** Randle's equivalent circuit, a circuit analogy of local interface impedance, where $C_{dl}$ is the double layer capacitance, $\rho_{ct}$ is the charge transfer resistance, and $z_D$ is the local diffusion impedance

where $\hat{j}_{tot}$ is the total current density. Using the total current density and the potential drop at the interface, we can define local interface impedance as

$$z_{intf} = \frac{\Delta\hat{\phi}}{\hat{j}_{tot}} = \left(i\omega C_{dl} + \left(\rho_{ct} + z_D\right)^{-1}\right)^{-1} \quad (14)$$

where $z_{intf}$ is the local interface impedance. The local interface impedance can be represented by the Randle's equivalent circuit, which has $C_{dl}$ in parallel with a series of $\rho_{ct}$ and $z_D$, as shown in Figure 2. As the model contains the parallel contribution of $C_{dl}$ and $\rho_{ct}$, a resistive-capacitance (RC) characteristic frequency naturally arises, $\omega_{RC} = \left(\rho_{ct}C_{dl}\right)^{-1}$, around which relative magnitudes of $\hat{j}_{dl}$ and $\hat{j}_{intc}$ are flipped.

The local interface impedance can be scaled by $\rho_{ct}$ to give its dimensionless form.

$$\begin{aligned}\tilde{z}_{intf} = \frac{z_{intf}}{\rho_{ct}} &= \left(i\left(\omega_D\rho_{ct}C_{dl}\right)\tilde{\omega} + \left(1 + \frac{\rho_D}{\rho_{ct}}\tilde{z}_D\right)^{-1}\right)^{-1} \\ &= \left(i\left(\tilde{\omega}/\tilde{\omega}_{RC/D}\right) + \left(1 + \tilde{\rho}_{D/ct}\tilde{z}_D\right)^{-1}\right)^{-1}\end{aligned} \quad (15)$$

where $\tilde{\omega}_{RC/D} = \omega_{RC}/\omega_D$ is the characteristic frequency ratio, and $\tilde{\rho}_{D/ct} = \rho_D/\rho_{ct}$ is the dimensionless BD impedance coefficient. The two dimensionless parameters, $\tilde{\omega}_{RC/D}$ and $\tilde{\rho}_{D/ct}$, determine the behavior of the local interface impedance. $\tilde{\omega}_{RC/D}$ is a measure of the separation of the two characteristic frequencies, $\omega_{RC}$ and $\omega_D$. For most battery electrodes, $\tilde{\omega}_{RC/D}$ is a large number, and the local interface impedance leads to well-separated RC and BD elements in the overall impedance spectra; the RC element ideally draws a semicircle at high frequencies with its summit at $\omega_{RC}$, and the BD element draws a hockey-stick-like curve at low frequencies with its kink around $\omega_D$ in the Nyquist plot. Relative magnitudes of the two elements are determined by $\tilde{\rho}_{D/ct}$.

*Overall impedance response* − While local impedance response has been considered to this point, impedance spectroscopy measures the integrative response of an entire electrode that may involve uneven local impedance response on its surface. Thus overall electrode impedance is defined with a total current which could be obtained by integrating $j_{tot}$ over the entire electroactive surface. When a thin film electrode has a uniform thickness, $j_{tot}$ is even over the entire surface and the integration results in a trivial scaling of $j_{tot}$ by the total area. On the other hand, although each isotropic particle has even $j_{tot}$ on its surface, a nanoparticle electrode has non-uniform $j_{tot}$ across particles due to the heterogeneity in particle size. In such cases, a particle size, which is the solid-state diffusion length, $l$, in the diffusion model, is considered as a realization of a continuous random variable with a certain probability density function (PDF). Then the integration could be performed with respect to $l$, in which $j_{tot}$ is weighted by a product of particle population and surface area.

$$\hat{J} = \int \hat{j}_{tot}(l) dA = N_{tot} \int_0^\infty \Pr_L(l) \bar{a}_p(l) \hat{j}_{tot}(l) dl \qquad (16)$$

where $J$ is the total current, $A$ is the electroactive surface area, $N_{tot}$ is the total number of particles, and $\Pr_L$ is the PDF of the solid-state diffusion length, a random variable, $L$. $\bar{a}_p(l)$ is the average surface area of a single particle with $L = l$; $\bar{a}_p(l)$ is $2\bar{a}_x$ for planar particles, $2\pi \bar{H} l$ for cylindrical particles, and $4\pi l^2$ for spherical particles, respectively, where $\bar{a}_x$ is the average sidewall area of the planar particles and $\bar{H}$ is the average height of the cylindrical particles.

Also, an impedance measurement inevitably involves interferences from cell connections as well as transports of ions in the electrolyte phase, whose contribution could be modeled with a resistor in the typical frequency window of an impedance measurement. Therefore, the overall impedance of a nanoparticle electrode can be written as follows.

$$Z = R_{ext} + \frac{\Delta \hat{\phi}}{\hat{J}} = R_{ext} + \frac{\Delta \hat{\phi}}{N_{tot} \int_0^\infty \Pr_L(l) \bar{a}_p(l) \hat{j}_{tot}(l) dl}$$
$$= R_{ext} + \left( N_{tot} \int_0^\infty \Pr_L(l) \bar{a}_p(l) z_{intf}^{-1}(l) dl \right)^{-1} \qquad (17)$$

where $R_{ext}$ is the external resistor that represents the contribution from cell connections and transports in electrolyte phase.

When $L$ has a narrow enough distribution, its PDF can be approximated by a Dirac delta function, which makes the integration trivial. This approximation is equivalent to assuming an identical particle size in a nanoparticle electrode. Under such a condition, the overall impedance becomes

$$Z = R_{ext} + \left( N_{tot} \int_0^\infty \delta(l - \bar{L}) \bar{a}_p(l) z_{intf}^{-1}(l) dl \right)^{-1} = R_{ext} + \frac{z_{intf}(\bar{L})}{A_{tot}} \tag{18}$$

where $\delta$ is the Dirac delta function, $\bar{L}$ is the average solid-state diffusion length, and $A_{tot}$ is the total surface area. The overall impedance of a uniform thin film electrode has the same expression, having one-dimensional BD impedance in $z_{intf}$. The argument here does not account for gradients in potential or ion concentrations that may develop along the electrode thickness in a porous nanoparticle electrode. While the gradients are negligible in thin nanoparticle electrodes, detailed elaborations regarding their effects on impedance of battery electrodes can be found in reference 15.

The overall impedance in Equation (17) can be reduced to its dimensionless form, defining the dimensionless overall impedance, $\tilde{Z} = A_{tot} Z / \rho_{ct}$, and the dimensionless solid-state diffusion length, $\tilde{l} = l / \bar{L}$, along with the dimensionless variables defined previously. The frequency and the RC characteristic frequency are now scaled by $\omega_D(\bar{L}) = D_{ch} / \bar{L}^2$, and their dimensionless forms become $\tilde{\omega} = \omega / \omega_D(\bar{L})$ and $\tilde{\omega}_{RC/D} = \omega_{RC} / \omega_D(\bar{L})$. Expanding the local interface impedance, $\tilde{z}_{intf}$, the overall impedance becomes

$$\begin{aligned}
\tilde{Z} &= \frac{A_{tot} Z}{\rho_{ct}} = \frac{A_{tot} R_{ext}}{\rho_{ct}} + \left( \int_0^\infty \Pr_{\tilde{L}}(\tilde{l}) \tilde{a}_p(\tilde{l}) \tilde{z}_{intf}^{-1}(\tilde{l}) d\tilde{l} \right)^{-1} \\
&= \tilde{R}_{ext} + \left( i(\tilde{\omega} / \tilde{\omega}_{RC/D}) + \int_0^\infty \Pr_{\tilde{L}}(\tilde{l}) \tilde{a}_p(\tilde{l}) \left(1 + \tilde{\rho}_{D/ct}(\tilde{l}) \tilde{z}_D(\tilde{l}) \right)^{-1} d\tilde{l} \right)^{-1}
\end{aligned} \tag{19}$$

Here, $\tilde{R}_{ext} = A_{tot} R_{ext} / \rho_{ct}$ is the dimensionless external resistance, and it indicates the relative magnitude of the external resistance with respect to the charge transfer resistance. $\Pr_{\tilde{L}}$ is the PDF of the dimensionless solid-state diffusion length, a random variable, $\tilde{L} = L / \bar{L}$. When a lognormal PDF is used for $\Pr_{\tilde{L}}$, it can be solely described by the dimensionless standard deviation, $\tilde{\sigma}$,

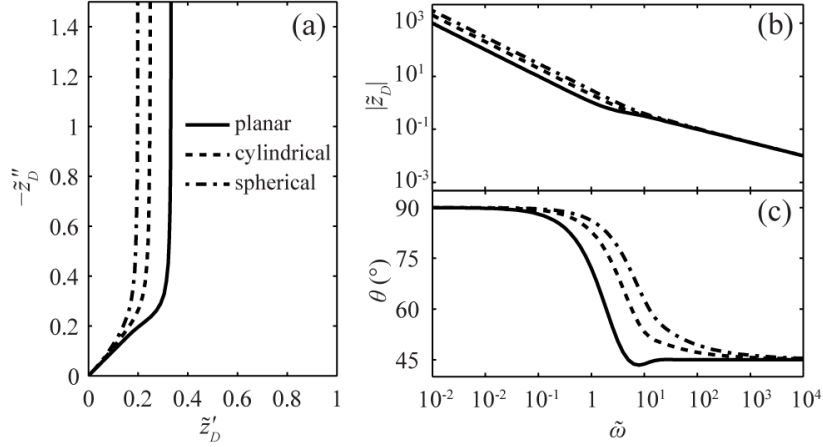

**Figure 3.** Dimensionless local bounded diffusion impedance for different particle geometries: (a) Nyquist plot, (b) magnitude plot, and (c) phase plot

which is a measure of the heterogeneity in particle size. $\tilde{\bar{a}}_p(\tilde{l}) = \bar{a}_p(l)/\bar{a}_p(\bar{L}) = \tilde{l}^{n-1}$ is the dimensionless average surface area of a single particle with $\tilde{L} = \tilde{l}$, and it gives different weighting distributions on the local impedance in the integral, depending on the particle geometry.

### Results

*Effect of nanoparticle geometry* –To isolate the effect of solid-state diffusion geometry on diffusion impedance, we first look at the local BD impedance, $\tilde{z}_D$. Figure 3 shows $\tilde{z}_D$ for the model diffusion geometries in various formats. For all the diffusion geometries, a clear transition is observed near $\tilde{\omega} \approx 1$, around which the diffusion penetration depth reaches the impermeable current collector of a thin film electrode or the symmetric center of a particle. At $\tilde{\omega} \gg 1$, a Warburg regime is defined, where $\tilde{z}_D$ asymptotically approaches the original Warburg behavior in its high frequency limit. In this regime, the penetration depth is shorter than the diffusion length, $l$, and the bounded diffusion behaves much like semi-infinite diffusion. On the other hand, at $\tilde{\omega} \ll 1$, a capacitive regime is defined, where $\tilde{z}_D$ has a capacitive behavior, drawing a vertical line in the Nyquist plot and approaching $90°$ in the phase angle plot. In this regime, the penetration depth exceeds $l$, and the perturbation leads to effectively filling up and emptying the active material, much like a capacitor.

In both regimes, $\tilde{z}_D$ has different behavior depending on diffusion geometry, as denoted by its asymptotic approximations in high and low frequencies, $\tilde{z}_{D\infty}$ and $\tilde{z}_{D0}$, respectively. For the planar diffusion geometry, $\tilde{z}_{D\infty}$ has the form of the original Warburg impedance, and $\tilde{z}_D$ follows the original Warburg behavior in most of the Warburg regime. On the other hand, for the curved diffusion geometries, the cylindrical and spherical models, $\tilde{z}_{D\infty}$ retains an extra imaginary term in addition to the original Warburg formula. This implies that for the curved diffusion geometries, $\tilde{z}_D$ in the Warburg regime can be approximated by a series circuit of the original Warburg impedance and a capacitance, $\tilde{C}_{D\infty} = 2/(n-1)$. Correspondingly in this regime, the $\tilde{z}_D$ curves have positive deviation in phase angle, or capacitive deviation, from the original Warburg behavior. The deviation reflects the variation in the flux-normal area with respect to the distance from the active material/electrolyte interface; it is larger for the spherical geometry than for the cylindrical.

On the other hand, $\tilde{z}_{D0}$ indicates that $\tilde{z}_D$ in the capacitive regime can be analogized to a series circuit of a resistance, $\tilde{\rho}_{D0} = 1/(n+2)$, and a capacitance, $\tilde{C}_{D0} = 1/n$, whose values differ for the different diffusion geometries. The differences in this regime arise primarily from different aspect ratios of the model geometries. $\tilde{\rho}_{D0}$ is smaller for the more-curved geometry; that is, it is smaller for the spherical geometry than for the cylindrical, and it is smaller for the cylindrical geometry than for the planar geometry. This difference exists because it is easier to diffuse throughout the entire domain with a higher surface to volume ratio, given an identical diffusion length and driving force on the surface. $\tilde{C}_{D0}$ is also smaller for the more-curved geometry, because it has smaller volume to accommodate ions. Correspondingly, in the Nyquist plot, a curve for the more-curved geometry is shunted at a smaller real projection, and has a higher imaginary projection given an identical frequency.

While the transition from the Warburg regime to the capacitive regime takes place near $\tilde{\omega} \approx 1$ regardless of the diffusion geometry, its relative location differs depending on the diffusion geometry. Since the penetration depth propagates deeper in the more-curved geometry, given an identical frequency, the transition in $\tilde{z}_D$ occurs at relatively higher frequencies. Therefore, in the absence of the effect of particle size distribution, $\omega_D$ as well as $D_{ch}$ are significantly overestimated, if a simple one-dimensional $\tilde{z}_D$ model is employed in interpreting impedance spectra of a curved-particle electrode. The overestimation is expected to be larger when the actual particles have the more-curved diffusion geometry.

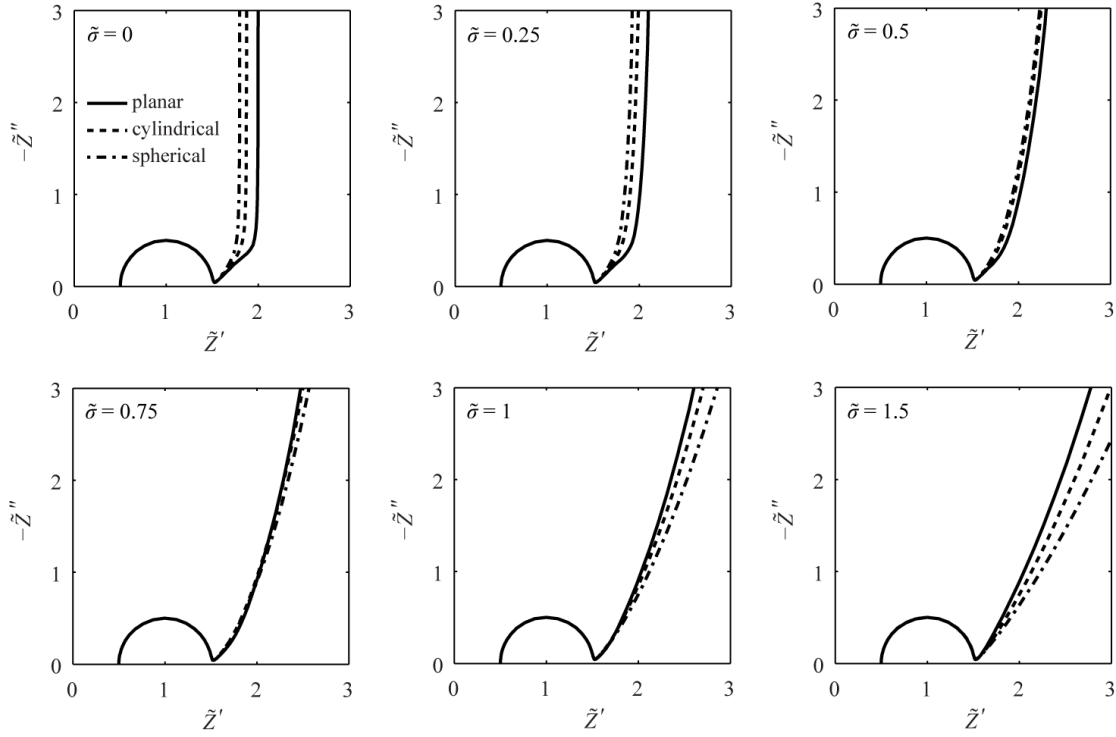

**Figure 4.** Dimensionless overall impedance of nanoparticle electrodes with different particle geometries, varying the standard deviation of particle size distribution (in Nyquist plots)

*Effect of particle size distribution* – The effect of heterogeneity in solid-state diffusion length, $\tilde{l}$, in particles appears in the overall impedance, $\tilde{Z}$, given by Equation (19). The dimensionless parameters required in the model are taken from literature values [15, 27], and a lognormal PDF is employed to describe the distribution in $\tilde{l}$. Figure 4 and 5 show $\tilde{Z}$ for the model electrode configurations, varying the standard deviation, $\tilde{\sigma}$. Figure 4 shows their behaviors in the Nyquist plot, and Figure 5 shows their behaviors in magnitude and phase angle plots. Since they involve the Randle-type local interface model (Figure 2) and well-separated characteristic frequencies ($\tilde{\omega}_{RC} \gg 1$), all impedance spectra in Figure 4 and 5 have well-distinguished RC and BD elements at high and low frequencies, respectively.

Regardless of the particle geometry, the heterogeneity in particle size makes the BD element of the overall impedance, or the overall BD impedance, deviate from the pristine behavior of the local BD impedance. When $\tilde{\sigma} = 0$ for an electrode with an identical particle size, the solid-state diffusion in its particles has a single $\omega_D$ without any dispersion, and the overall BD impedance keeps the pristine transition behavior observed in the local BD impedance near $\tilde{\omega} \approx 1$. On the other hand, when an electrode has heterogeneous particles sizes, the diffusion has a dispersion in $\omega_D$, and the transition in the overall BD impedance spreads over a wider frequency range. $\tilde{z}_D$ of a smaller particle ($\tilde{l} < 1$) transitions from the Warburg regime to the capacitive regime around relatively higher $\omega_D$, compared to that of an average-size particle. Their contribution gives positive deviation in phase angle, or capacitive deviation, in the Warburg regime of the overall BD impedance. On the other hand, $\tilde{z}_D$ of a larger particle ($\tilde{l} > 1$) transitions around relatively lower $\omega_D$, and gives negative deviation in phase angle, or resistive deviation, in the capacitive regime. These two kinds of contribution are combined to make the transition in the overall BD impedance smoother. When $\tilde{\sigma} \geq 0.5$, the transition become quite blurred, and the Warburg and the capacitive regimes are not distinguishable. It shows that the particle size distribution may be another reason for the constant phase element (CPE)-like behavior in the lower frequency range in diffusion impedance of battery electrodes [13, 28].

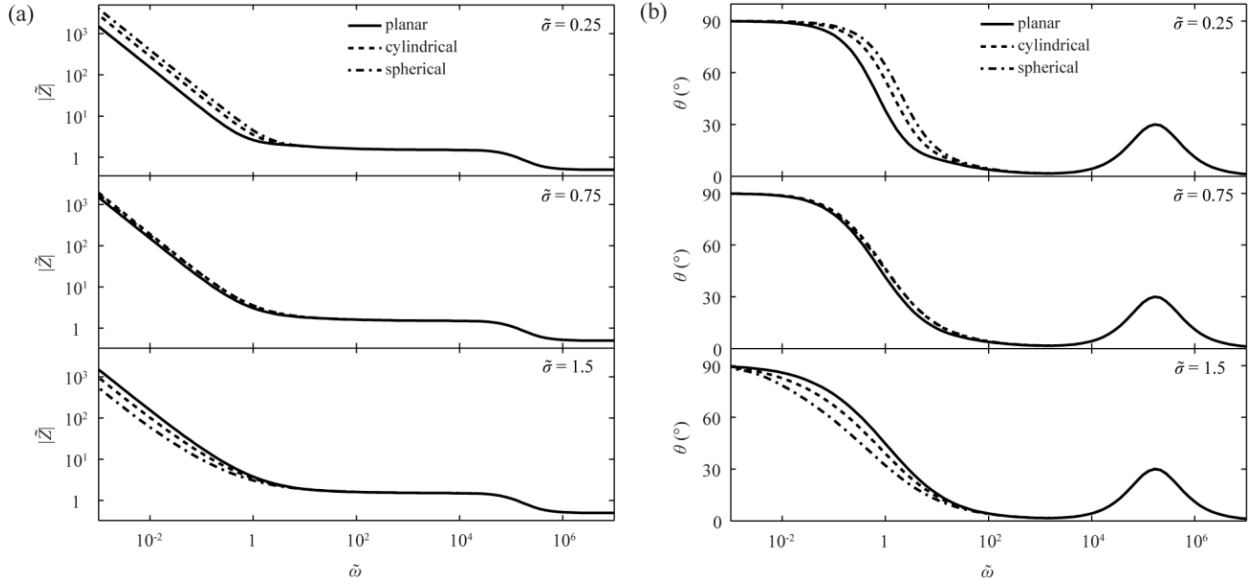

**Figure 5.** Dimensionless overall impedance of nanoparticle electrodes with different particle geometries, varying the standard deviation of particle size distribution: (a) magnitude plots and (b) phase plots

The deviation due to the heterogeneity in particle size is more significant for an electrode with the more-curved particles; it is larger for the spherical-particle electrode than for the cylindrical-particle electrode, and it is larger for the cylindrical-particle electrode than the planar-particle electrode. The difference in the extent of the deviation changes the effect of particle geometry on the BD impedance. The overall BD impedance curves near $\tilde{\sigma} = 0$ maintain the pristine trend with respect to particle geometry that is observed in the local BD impedance; the hockey-stick-like curves in the Nyquist plot are shunted more to the left as the particle geometry becomes more-curved. Also the transition takes place over a higher frequency range as the particle geometry becomes more-curved. However, when $0.5 \leq \tilde{\sigma} \leq 0.75$, the curves show very similar behaviors regardless of the particle geometry. When $\tilde{\sigma} \geq 1.0$, the trend is switched; overall BD impedance has a smaller overall envelope in the Nyquist plot, and transits over a wider and lower frequency range, as the particle geometry is more-curved. A larger extent of deviation is observed for the more-curved particle geometry because $\tilde{\bar{a}}_p$ places relatively heavier weighting on the local response of large particles ($\tilde{l} > 1$).

The monotonic correlation between $\tilde{\sigma}$ and the extent of deviation implies that it is possible to estimate $\tilde{\sigma}$ from the BD impedance spectra, once the particle geometry and an appropriate form of the PDF are known. This estimation would be a good supplement for the existing combination of an electron microscopy and an automatic image analyzer, since an impedance measurement probes the entire electrode, rather than a local image, without opening the cell.

Since heterogeneity in particle size makes the transition take place in a lower frequency range overall, it results in underestimation of $\omega_D(\bar{L})$ as well as $D_{ch}$, if a model of an identical particle size is employed in interpreting impedance spectra of an electrode that has heterogeneous particle sizes. The extent of underestimation would be larger when the distribution in particle size is more heterogeneous, and when the particle geometry is more-curved. When an electrode has a broad distribution in particle size ($\tilde{\sigma} \geq 0.5$), the model with an identical particle size would fail to match the experimental spectra, leading to apparently poor agreement for the diffusion impedance.

Unlike the previous discussion in the absence of particle size distribution, it leads to different results depending on the distribution in particle size, if a one-dimensional $\tilde{z}_D$ model is employed to interpret impedance spectra of a curved-particle electrode. When the electrode has a narrow particle size distribution ($\tilde{\sigma} \leq 0.25$), overlooking the particle curvature overestimates $D_{ch}$, similarly to the absence of particle size distribution. On the other hand, when the distribution is intermediate ($0.5 \leq \tilde{\sigma} \leq 0.75$), it results in negligible disagreement in $D_{ch}$, because the overall BD impedance is insensitive to particle geometry. When the distribution is broad ($\tilde{\sigma} \geq 1.0$), reasonable agreement results only if $\tilde{\sigma}$ is left as a fit parameter; in such cases, $\tilde{\sigma}$ is overestimated by overlooking the particle curvature, but the direction of misestimation in $D_{ch}$ is ambiguous, depending on the actual particle geometry and size distribution.

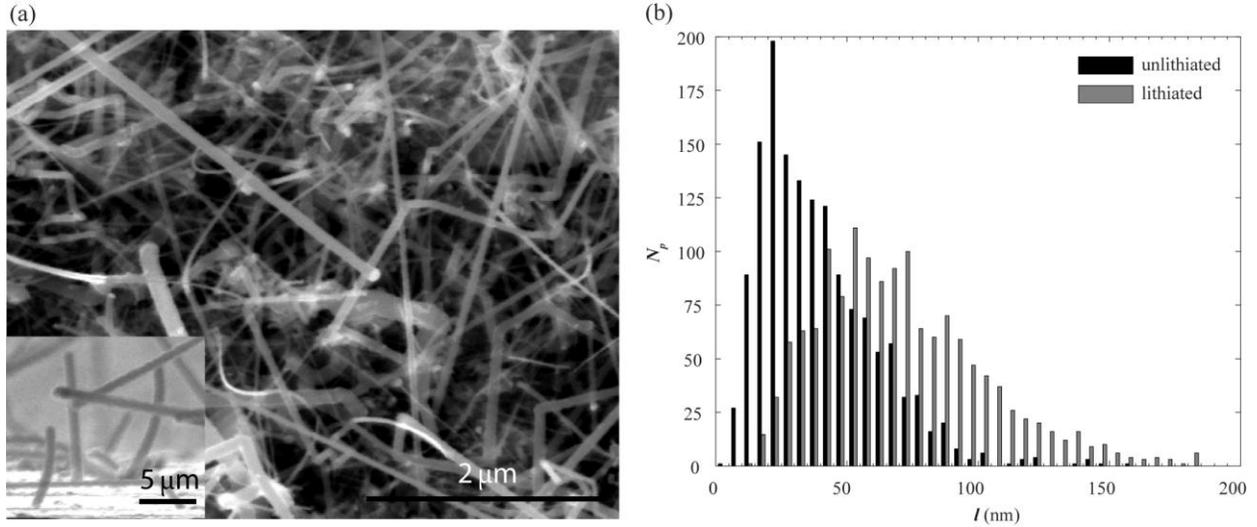

**Figure 6.** (a) SEM image of pristine Si nanowire electrode and (b) radius distributions of Si nanowires, adapted by permission from Macmillan Publishers Ltd: *Nature Nanotechnology* (reference 34), copyright 2007

**Application and Discussion**

One of the systems where this model could be well-applied is a lithium (Li) ion battery electrode made of silicon (Si) nanowires. Although pristine Si is one of the most common semiconducting materials, the assumption of fast electron mobility is valid when Si is doped with Li for a wide range of concentration [29, 30]. Moreover, Si nanowires after an initial lithiation remain amorphous for a wide range of Li concentration, and the transport properties of the charge carriers are isotropic [31-33]. One of the most distinguished features of Si nanowires is their well-defined cylindrical geometry with a high length-to-radius ratio. Their well-defined geometry combined with the isotropic transport properties makes the cylindrical model of BD impedance applicable to impedance spectra of a Si nanowire electrode.

Impedance behavior of a Si nanowire porous electrode has been investigated by Ruffo *et al.*, in 2009, and the raw data were kindly provided to us by Professor Yi Cui [8]. The details of experimental procedures and characterization methods are reported in their original papers [8, 34]. The electrode configuration is shown in Figure 6, along with typical radius distributions of the nanowires. The distributions have lognormal-like forms with mean radii of 44.5 nm and 70.5 nm, and standard deviations of 22.5 nm and 32.0 nm, for nanowires that are fully delithiated and lithiated, respectively. The thickness of the electrode is around 30 $\mu$m, and it turns out to be thin enough that the effect of concentration and potential gradients along the thickness is negligible in its impedance spectra, which would have made the impedance curve skewed [15, 16, 35].

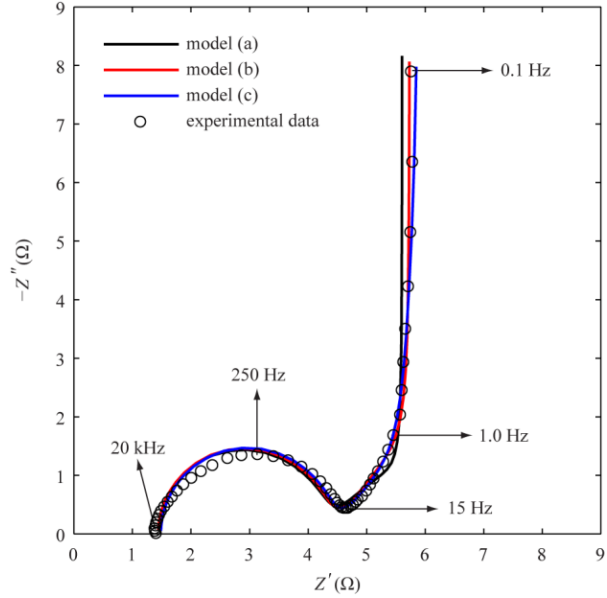

**Figure 7.** Representative impedance spectra of Si nanowire electrode and model fit curves at $c_{Li}$ = 954 mAh/g

A representative impedance spectrum of the Si nanowire electrode at Li concentration of 954 mAh/g is shown in Figure 7. It shows the typical behavior of electrode impedance with the Randle's local interface model and well-separated characteristic frequencies ($\omega_{RC} \gg \omega_D$). The high frequency (20 kHz) intercept on the real axis corresponds to the value of $R_{ext}$. The semicircle at high frequencies (10 kHz ~ 15 Hz) is the RC element which represents the gradual transition from double layer-dominating response to intercalation-dominating response around $\omega_{RC}$. At low frequencies (below 15 Hz), the BD element appears, which shows a transition from the Warburg regime (15 Hz ~ 1 Hz) to the capacitive regime below 1 Hz.

To study the effect of including the actual nanowire geometry and the radius distribution in modeling, three different models were employed in fitting the experimental data; model (a) has planar particles with identical thickness, model (b) has cylindrical particles with identical radii, and model (c) has cylindrical particles with distributed radii. While model (a) ignores the primary curvature of the cylindrical diffusion geometry in the nanowires, models (b) and (c), include the actual cylindrical geometry of the nanowires. Model (c) takes account of the radius distribution of the nanowires, whereas model (b) neglects the distribution and considers the radii identical. Complex nonlinear least squares (CNLS) regression was performed using a MATLAB routine based on the Levenberg-Marquardt algorithm that minimizes the summed squares of real and imaginary relative residuals:

$$\Delta_{re,k} = \frac{Z'_k - Z'(\omega_k)}{|Z_k|}, \quad \Delta_{im,k} = \frac{Z''_k - Z''(\omega_k)}{|Z_k|} \tag{20}$$

where $\omega_k$ is the $k$ th frequency in the impedance measurement, $Z_k$ is the impedance measured at $\omega_k$, and $Z$ is the overall impedance model. The data and fitted curves are shown in Figure 7, and the obtained parameters are shown in Table 2.

The effect of including the actual nanowire geometry in modeling can be studied by comparing the fit curves of model (a) and (b). Model (b), which involves the cylindrical diffusion geometry in the nanowires, has a better agreement with the experimental data than model (a), particularly for the BD element. The improvement in goodness-of-fit is quantified by a sum of relative residual squares for the BD element ($\omega < 15$ Hz), which is denoted as $\Sigma$. $\Sigma$ is 0.0053 in model (a), but shrinks to 0.0017 in model (b) with the same number of fit parameters. The relative residuals for the BD element are plotted against $\omega$ in Figure 8. It is shown that the locally correlated residuals are reduced by accounting for the actual cylindrical geometry of the nanowires in model (b). The improvement in goodness-of-fit partially advocates employing the cylindrical diffusion model in interpreting impedance spectra of a Si nanowire electrode.

The chemical diffusivities, $D_{ch}$, obtained by employing model (a) and model (b) are 3.57 × $10^{-11}$ and 1.25 × $10^{-11}$ cm$^2$/sec, respectively. Overlooking the primary curvature of the cylindrical nanowire geometry in model (a) results in overestimation of $D_{ch}$ by more than two and half times, compared to that obtained considering the actual geometry in model (b). This indicates that to obtain an accurate diffusivity from impedance spectra, it is necessary to employ an appropriate diffusion geometry rather than a simple one-dimensional diffusion model. For an electrode with more-curved particles such as sphere-like particles, the overestimation would be even more significant. On the other hand, the pseudo film model suggested by Ruffo *et al.* has the one-dimensional BD impedance formula with a single diffusion length, as model (a) does. However, their calculation estimates the diffusivities between 1.5 and $3.0 \times 10^{-10}$ cm$^2$/sec, about an order larger compared to that obtained from model (a), probably because a different length scale was used.

Table 2. Parameters estimated by fitting different versions of the model to impedance spectra at $c_{Li}$ = 954 mAh/g

|  | Model (a) | Model (b) | Model (c) |
| --- | --- | --- | --- |
| $D_{ch}$ (× $10^{-11}$ cm$^2$/sec) | 3.57 | 1.25 | 1.45 |
| $\Sigma$ | 0.0053 | 0.0017 | 5.4 × $10^{-4}$ |
| $\tilde{\sigma}$ | -- | -- | 0.20 |

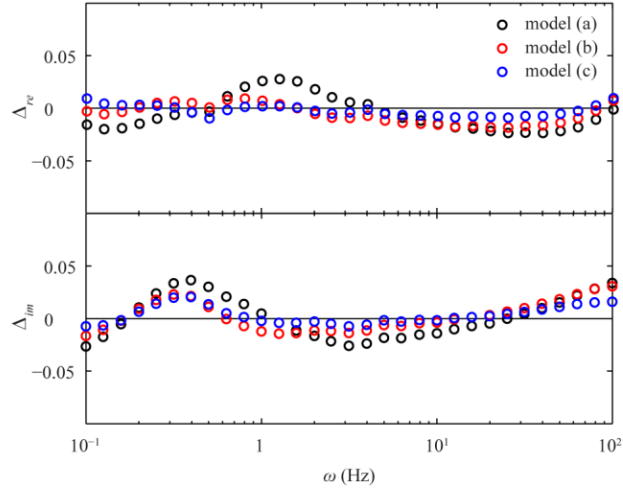

**Figure 8.** Relative residuals for the bounded diffusion impedance at $c_{Li}$ = 954 mAh/g

Additional consideration on the radius distribution in model (c) employs a lognormal PDF, leaving the standard deviation, $\tilde{\sigma}$, as an additional fit parameter. The fit of model (c) now captures the deviation of the diffusion element from the pristine shape of local BD impedance, which is a generic feature in impedance spectra of a battery electrode that has heterogeneous particle sizes. Simultaneously, $\Sigma$ decreases to $5.4 \times 10^{-4}$, and the locally correlated residuals are reduced by using model (c) as shown in Figure 8. $\tilde{\sigma}$ is estimated to be 0.20, which is smaller than but still comparable to those of the typical radius distributions.

Given the extremely good fit of the impedance spectrum with model (c), the inferred particle size distribution from the impedance model may be more accurate than that obtained by the typical, time-consuming method of direct image analysis. This suggests the tantalizing possibility of non-invasive "impedance imaging" of battery electrodes *in situ* by the simple application of small electrical signals, using the diffusion-impedance model to solve an inverse problem for the particle size distribution. Such changes in particle thickness could be used, for example, to detect the state of charge (due to volume change upon lithiation) or gradual degradation over many cycles, e.g. due to the formation of solid electrolyte interphase. Mathematically, the inversion can be performed by choosing a functional form (such as log-normal) for the size distribution and solving for the best-fit parameters, as we have done here. It is also possible to view the inverse problem as a first-kind Fredholm integral equation for the unknown size distribution function, which can be solved by Laplace or Mellin transforms, as has been done for various inverse problems in statistical mechanics [36].

Using model (c) in the regression, $D_{ch}$ is estimated to be $1.45 \times 10^{-11}$ cm$^2$/sec. Comparing $D_{ch}$ values obtained from model (b) and (c), it is found that overlooking the radius distribution in model (b) results in underestimation of $D_{ch}$. In this particular case of a Si nanowire electrode with estimated $\tilde{\sigma}$ of 0.20, the underestimation is about 16 %. In general, the extent of the underestimation would be larger when the distribution is more heterogeneous or when the particle geometry is more-curved.

Impedance spectra of the Si nanowire electrode at various Li concentrations are shown in Figure 9. For the intermediate concentrations, the spectra were fitted using model (c), taking into account the cylindrical diffusion geometry as well as the radius distribution of the nanowires. The fit curves are also plotted in Figure 9, and $D_{ch}$ values obtained from the regression are shown in Table 3. $D_{ch}$ is estimated in the range of $1.18 \sim 2.01 \times 10^{-11}$ cm$^2$/sec, depending on the Li concentration. These values agree well with the diffusivities measured by N. Dimov *et al.* for a Si powder electrode, which are 1.7 and $6.4 \times 10^{-11}$ cm$^2$/sec at Li concentrations of 800 and 1200 mAh/g, respectively [37]. Reported values of the Li diffusivity in amorphous Si are inconsistent and spread over a wide range, varying from the order of $10^{-11}$ to $10^{-13}$ cm$^2$/sec at room temperature [38-40]. The excellent fit of the impedance spectrum using the known particle geometry and size distribution suggests that our value is among the most accurate in the literature.

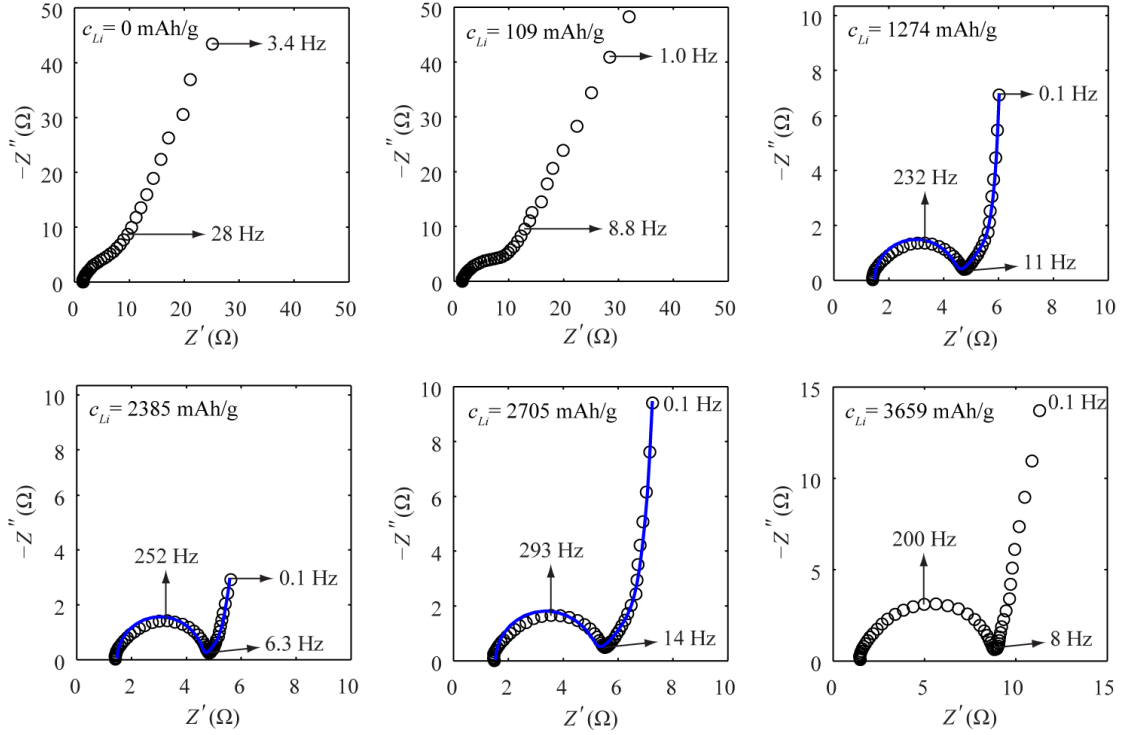

**Figure 9.** Impedance spectra and fit curves at different Li concentrations during the second cycle

Limitations of the diffusion model, however, can be found at both of low and high Li concentrations. At the low Li concentrations, it is difficult to identify either RC or BD elements in the spectra. The two elements seem overlapped, and the behavior at low frequencies is different from the typical BD impedance observed at the intermediate Li concentrations. The model is not able to interpret these features. A possible explanation is that the assumption regarding fast electron transport is not valid at low Li concentrations, as the electron mobility becomes orders of magnitude smaller than at higher Li concentrations [29]. Although it is beyond the scope in this article, an accurate interpretation of the impedance spectra at the low Li concentrations may involve solving simultaneous transport of Li ions and electrons from two different kinds of interfaces: radial diffusion of ions from the electrolyte and axial diffusion of electrons from the current collector. On the other hand, when Li concentration approaches the full capacity, cycled Si nanowires encounter rapid phase transformation from the amorphous phase to the $Li_{15}Si_4$ crystalline phase [32, 33]. When it comes to the crystalline phase at high Li concentrations, the assumption regarding isotropic transport properties is not valid any more, and our model would need to be modified to account for the onset of crystal anisotropy, as well as possibly the dynamics of the phase transformation.

**Table 3. The chemical diffusivities of Li ion in amorphous Si nanowire at different Li concentrations during the second cycle**

| Cycle status | $c_{Li}$ (mAh/g) | $D_{ch}$ ($\times 10^{-11}$ cm$^2$/sec) |
|---|---|---|
| 2 (charging) | 954 | 1.45 |
| 2 (discharge) | 1274 | 1.29 |
| 2 (charging) | 2385 | 1.18 |
| 2 (discharge) | 2705 | 2.01 |

**Conclusion**

Modern battery electrodes have nanoparticles of active material which have various shapes and heterogeneous sizes. In impedance spectra of such electrodes, the responses at low frequencies correspond to the bounded diffusion of ions in the particles. While properties of the BD impedance are expected to essentially depend on the diffusion geometry and the diffusion length distribution in the nanoparticles, the effects of such configurational aspects have been largely overlooked. Commercial data-processing software products only contain the one-dimensional BD impedance model and the original Warburg impedance model, which are not able to account for these effects. In this study, an impedance model is proposed that accounts for curved diffusion geometries as well as the diffusion length distribution. Using this model, we have investigated the ways these configurational aspects affect interpretation of diffusion impedance spectra. The model also opens the possibility of conversely using impedance spectroscopy to diagnose battery electrodes in terms of the configuration-related status, *in-situ* during a test that requires many cycles.

Various versions of the model were then applied to experimental impedance data of a Si nanowire electrode. Comparing the regression results of the different versions, we are able to show the effect of including each of the cylindrical diffusion geometry and the heterogeneous radius distribution of the nanowires greatly improves the fit and leads to rather different, and presumably more accurate, values of the Li diffusivity. In this particular example, overlooking the curvature of the cylindrical diffusion geometry and applying a one-dimensional BD impedance model results in overestimation of the Li diffusivity by more than two and half times.

On the other hand, assuming an identical particle size results in underestimation of the diffusivity by 16 %, compared to one obtained including the radius distribution in the model. In general, the effects of including appropriate particle geometry and particle size distribution in modeling depend on the actual particle geometry and size distribution of an electrode. To accurately interpret diffusion impedance of a battery electrode, it is important to account for the nanoparticle geometry and the size distribution.

**Acknowledgment**

This work was partially supported by a grant from the Samsung-MIT Alliance and by a fellowship to JS from the Kwanjeong Educational Foundation. The authors thank Professor Yi Cui at Stanford University for providing the raw experimental data for silicon nanowire anodes.

**List of Symbols**

| | |
|---|---|
| $\bar{a}_p$ | average surface area of a single particle [cm$^2$] |
| $\bar{a}_x$ | average sidewall area of a planar particle [cm$^2$] |
| $\tilde{\bar{a}}_p$ | dimensionless surface area of a single particle ($\tilde{\bar{a}}_p = L^{n-1}$) |
| $A$ | electroactive surface area [cm$^2$] |
| $A_{tot}$ | total electroactive surface area [cm$^2$] |
| $A_W$ | Warburg coefficient [$\Omega$/s$^{1/2}$] |
| $c$ | concentration of ionic charge carrier [mol/cm$^3$] |
| $C_{dl}$ | local double layer capacitance [F/cm$^2$] |
| $\tilde{C}_{D\infty}$ | high-frequency-limit dimensionless extra capacitance of bounded diffusion |
| $\tilde{C}_{D0}$ | low-frequency-limit dimensionless capacitance of bounded diffusion |
| $D_{ch}$ | chemical diffusivity of positive charge carrier [cm$^2$/s] |
| $e$ | elementary electric charge [C] |
| $\bar{H}$ | average height of a cylindrical particle [cm] |
| $j_0$ | exchange current density [A/cm$^2$] |
| $j_{dl}$ | double layer charging current density [A/cm$^2$] |
| $j_{intc}$ | intercalation current density [A/cm$^2$] |
| $j_{tot}$ | total current density [A/cm$^2$] |
| $J$ | total current [A] |
| $k$ | Boltzmann's constant [eV/K] |
| $l$ | solid-state diffusion length [cm] |
| $\tilde{l}$ | dimensionless solid-state diffusion length ($\tilde{l} = l/\bar{L}$) |
| $L$ | solid-state diffusion length, a random variable [cm] |
| $\bar{L}$ | average solid-state diffusion length [cm] |
| $\tilde{L}$ | dimensionless solid-state diffusion length, a random variable ($\tilde{L} = L/\bar{L}$) |
| $n$ | dimension number |
| $N_{tot}$ | total number of particles |
| $\Pr_L$ | probability density function of a random variable, $L$ [cm$^{-1}$] |
| $\Pr_{\tilde{L}}$ | probability density function of a random variable, $\tilde{L}$ |
| $q_{dl}$ | double layer charge [C/cm$^2$] |
| $R_{ext}$ | external resistance contribution [Ω] |
| $\tilde{R}_{ext}$ | dimensionless external resistance contribution ($\tilde{R}_{ext} = A_{tot}R_{ext}/\rho_{ct}$) |
| $t$ | time [s] |
| $T$ | temperature [K] |
| $x$ | spatial variable [cm] |
| $X$ | arbitrary variable |
| $X_{ref}$ | arbitrary variable at reference state |
| $\hat{X}$ | Fourier coefficient of perturbation in $X$ |
| $z_D$ | local diffusion impedance [Ωcm$^2$] |
| $\tilde{z}_D$ | dimensionless local diffusion impedance ($\tilde{z}_D = z_D/\rho_D$) |
| $\tilde{z}_{D0}$ | low-frequency-limit of dimensionless local diffusion impedance [Ωcm$^2$] |
| $\tilde{z}_{D\infty}$ | high-frequency-limit of dimensionless local diffusion impedance [Ωcm$^2$] |
| $z_{intf}$ | local interface impedance [Ωcm$^2$] |
| $\tilde{z}_{intf}$ | dimensionless local interface impedance ($\tilde{z}_{intf} = z_{intf}/\rho_{ct}$) |
| $Z_W$ | original Warburg impedance [Ω] |
| $Z$ | overall impedance [Ω] |
| $Z_k$ | measured impedance at frequency $\omega_k$ [Ω] |

| | | |
|---|---|---|
| $\tilde{Z}$ | | dimensionless overall impedance ($\tilde{Z} = A_{tot}Z/\rho_{ct}$) |
| | *Greeks* | |
| $\alpha$ | | symmetric coefficient |
| $\delta$ | | Dirac delta function [cm$^{-1}$] |
| $\Delta_{re,k}$ | | real relative residual at frequency $\omega_k$ |
| $\Delta_{im,k}$ | | imaginary relative residual at frequency $\omega_k$ |
| $\Delta\phi$ | | potential drop across active material/electrolyte interface [V] |
| $\Delta\phi_{eq}$ | | equilibrium potential of intercalation reaction [V] |
| $\rho_{ct}$ | | local charge transfer resistance [$\Omega$cm$^2$] ($\rho_{ct} = kT/j_0 e$) |
| $\rho_D$ | | BD impedance coefficient [$\Omega$cm$^2$] |
| $\tilde{\rho}_{D/ct}$ | | dimensionless BD impedance coefficient ($\tilde{\rho}_{D/ct} = \rho_D/\rho_{ct}$) |
| $\tilde{\rho}_{D0}$ | | low-frequency-limit resistance of bounded diffusion |
| $\sigma$ | | standard deviation [cm] |
| $\tilde{\sigma}$ | | dimensionless standard deviation ($\tilde{\sigma} = \sigma/\bar{L}$) |
| $\Sigma$ | | sum of relative residual squares for diffusion impedance |
| $\omega$ | | perturbation frequency [rad or Hz] |
| $\omega_D$ | | diffusion characteristic frequency [rad or Hz] ($\omega_{RC} = D_{ch}/L^2$) |
| $\omega_k$ | | $k$ th experimental frequency [rad or Hz] |
| $\omega_{RC}$ | | RC characteristic frequency [rad or Hz] ($\omega_{RC} = 1/\rho_{ct}C_{dl}$) |
| $\tilde{\omega}$ | | dimensionless perturbation frequency ($\tilde{\omega} = \omega/\omega_D$) |
| $\tilde{\omega}_{RC/D}$ | | characteristic frequency ratio ($\tilde{\omega}_{RC/D} = \omega_{RC}/\omega_D$) |
| $\eta$ | | surface overpotential [V] |